# Human Exposure to Radiofrequency Energy above 6 GHz: Review of Computational Dosimetry Studies

Akimasa Hirata[1,2], Sachiko Kodera[1], Kensuke Sasaki[3], Jose Gomez-Tames[1,2], Ilkka Laakso[4], Andrew Wood[5], Soichi Watanabe[3], Kenneth R. Foster[6]

1. Dept. of Electrical and Mechanical Engineering, Nagoya Institute of Technology, JAPAN
2. Center of Biomedical Physics and Information Technology, Nagoya Institute of Technology, Nagoya JAPAN
3. National Institute of Information and Communications Technology, Tokyo, JAPAN
4. Dept. of Electrical Engineering and Automation, Aalto University, Espoo, FINLAND
5. Swinburne University of Technology Melbourne, Australia
6. University of Pennsylvania, Philadelphia, USA

Corresponding author: Akimasa Hirata (e-mail: ahirata@nitech.ac.jp).

ABSTRACT:

International guidelines/standards for human protection from electromagnetic fields have been revised recently, especially for frequencies above 6 GHz where new wireless communication systems have been deployed. Above this frequency a new physical quantity "absorbed/epithelial power density" has been adopted as a dose metric. Then, the permissible level of external field strength/power density is derived for practical assessment. In addition, a new physical quantity, fluence or absorbed energy density, is introduced for protection from brief pulses (especially for shorter than 10 sec). These limits were explicitly designed to avoid excessive increases in tissue temperature, based on electromagnetic and thermal modeling studies but supported by experimental data where available. This paper reviews the studies on the computational modeling/dosimetry which are related to the revision of the guidelines/standards. The comparisons with experimental data as well as an analytic solution are also been presented. Future research needs and additional comments on the revision will also be mentioned.

## 1. Introduction

In 2019, a new wireless communications system named "5G" (5th generation) began to be deployed. The technology is presently defined to operate in three bands; "low" and "mid" which are similar to presently used cellphone bands, and a "high band" from 24 to 28 GHz that is higher than conventional wireless communication systems, e.g., 4G (2 GHz 3.5 GHz), wireless LAN (5.8 GHz) etc. This creates the need to assess human exposure to radiofrequency (RF) energy, both from uplink (handsets) and downlink (base station) and possibly the need to refine radiofrequency (RF) exposure limits and compliance assessment procedures at these higher frequencies.

In the previous version of the ICNIRP guidelines (1998) and the IEEE C95.1-2005 standard (2005) [1], the specific absorption rate (SAR) was the dosimetric or internal physical quantity for assessing exposure below 3 or 10 GHz, depending on the limit. At higher frequencies, the dosimetric measure changed to the incident power density, because of the shorter energy penetration depth in tissue. This introduced a discontinuity in the exposure limits across the transition frequency (Colombi *et al.*, 2015).

The recently updated limits, ICNIRP (2020) and IEEE C95.1-2019 (2019) have adopted a common "transition frequency" of 6 GHz. Below this frequency, the SAR remained the basic measure of internal exposure. Above this frequency a new metric for internal exposure has been adopted, "absorbed/epithelial power density". Both sets of limits were explicitly designed to avoid excessive increases in tissue temperature, based in large part on electromagnetic and thermal modeling studies but supported by experimental data where available (Foster *et al.*, 2016; Ziskin *et al.*, 2018; Hirata *et al.*, 2019).

This review summarizes recent advances in thermal and electromagnetic modeling of exposure at frequencies > 6 GHz. It comments also on the strengths and weaknesses of thermal models to predict the rise in temperature of RF-irradiated tissues.

## 2. Background

*A. Characteristics of Human Interaction with Biological Tissues*

---

[1] ICNIRP refers to its limits as "guidelines" while IEEE uses the term "standard". For conciseness the term "limit" is used to refer to either guideline/standard.

Microwave energy at frequencies > 6 GHz is absorbed close to the surface of the body. Table 1 summarizes the energy penetration depth and energy transmission coefficient, calculated from a simplified one-dimensional model (Foster *et al.*, 2018b) for tissue with dielectric properties of dry skin (Hasgall *et al.*, 2015).

Table 1. Power transmission coefficient and energy penetration depth into tissue

| Frequency, GHz | Power transmission coefficient into skin (Trans) | Energy penetration depth $L$ (mm) |
|---:|---:|---:|
| 6 | 0.47 | 3.7 |
| 10 | 0.49 | 1.9 |
| 30 | 0.54 | 0.43 |
| 100 | 0.70 | 0.18 |
| 300 | 0.84 | 0.14 |

* Adapted from Foster *et al.* (2018b), based on a uniform half plane of tissue with dielectric properties of dry skin. Power transmission coefficients are for plane waves normally incident on the surface. The energy penetration depth is defined as the distance beneath the surface at which the SAR has fallen to a factor of $1/e$ below that at the surface.

The dosimetry and thermal modeling problems extend over two different distance scales: exposure to RF energy is limited to tissues close to the body surface, whereas heat propagates into subcutaneous tissues and eventually is dissipated in the body core.

In the frequency range of present interest, the most relevant tissues for dosimetry are skin and cornea. Ziskin *et al.* (2018) reviewed the anatomy and electrical and thermal properties of skin that are relevant to assessments exposed to mm-wave (Table 2). The stratum corneum (outer layer) has comparatively low water content and varies in thickness in different parts of the body but is generally < 0.02 mm thick. The epidermis and dermis have much higher water content, and their combined thickness can exceed 2 mm. The deepest layer, the hypodermis, consists of subcutaneous fat and it also varies considerably in thickness among individuals.

Table 2. Thickness and water content of skin layers

| Skin layer | Thickness | | Water content (% weight) |
| --- | --- | --- | --- |
| | Typical (mm) | Range (mm) | |
| Stratum corneum (SC): | | | |
|   a) Thin skin | 0.015 | 0.01–0.02 | 0–10 |
|   b) Thick skin | 1.2 | 0.07–1.5 | 0–10 |
| Viable epidermis (epidermis minus SC) | 0.05 | 0.03–0.2 | 65–70 |
| Dermis | 1.4 | 0.5-2.2 | 60–70 |
| Hypodermis (fat): | | | |
|   a) Non-obese | 3 | 1.1-7.1 | 10–40 |
|   b) Obese | 20 | 10-30* | 10–40 |

From Table 1 in Ziskin *et al.* (2018). Fat thickness in obese individuals can exceed 30 mm.

*B. Dielectric Properties of Tissue*

For a comprehensive review of the dielectric properties of tissue; see Foster *et al.* (2018a) and Ziskin *et al.* (2018). Some of the most widely used data are from Gabriel *et al.* (1996), who measured the dielectric properties of 56 different human and animal tissues between 10 Hz and 20 GHz. The investigators fitted the data to Cole-Cole equations, and reported properties extrapolated to 100 GHz. With over 3700 citations (Google Scholar, September 2020), and a readily accessible online version, this dataset has become the *de facto* standard for dielectric properties for numerical dosimetry studies. It has been incorporated into commercial electromagnetic modeling programs.

Despite its widespread use, the Gabriel dataset has significant limitations. It was derived from a relatively small number of measurements on a limited number of tissue samples, many of them autopsy specimens or excised animal tissues, and the dataset may not accurately reflect the range of tissue properties *in vivo*. The highest measurement frequency was 20 GHz and entries at higher frequencies, up to 100 GHz, are extrapolated from lower frequencies, which is a potential source of error.

Tables 3 and 4 compare the real and imaginary parts of the complex relative permittivity ($\varepsilon'$ and $\varepsilon''$ respectively) and two derived quantities (the power transmission coefficient into tissue ($T_{tr}$) and energy penetration depth ($L$) of (rabbit) cornea at 30 and 100 GHz from the Gabriel dataset with measured data by Sasaki *et al.* (2015). The properties agree very well. It is not clear, however, how large the normal background variability in these properties will be. Skin is anatomically a multilayer structure whose layers vary significantly in water content and thickness, and consequently the bulk dielectric properties measured across the full thickness of skin will vary considerably (Ziskin *et al.*, 2018).

TABLE 3. Comparison of dielectric properties of rabbit cornea, 35-37 °C: Gabriel *et al.* (1996) (extrapolated) and Sasaki *et al.* (2015) (measured).

|  | Gabriel *et al.* (1996) | Sasaki *et al.* (2015) |
|---|---|---|
|  | 30 GHz | |
| $\varepsilon'$ | 20.9 | 19.5 |
| $\varepsilon''$ | 20.5 | 20.3 |
| $L$ (mm) | 0.39 | 0.38 |
| $T_{tr}$ | 0.50 | 0.50 |
|  | 100 GHz | |
| $\varepsilon'$ | 8.0 | 13.1 |
| $\varepsilon''$ | 10.3 | 7.0 |
| $L$ (mm) | 0.15 | 0.20 |
| $T_{tr}$ | 0.63 | 0.63 |

TABLE 4. Comparison of dielectric properties of skin, 35-37 °C: Gabriel *et al.* (1996) (extrapolated) and Sasaki *et al.* (2014) (measured).

|  | dry skin in Gabriel *et al.* (1996) | Dermis in Sasaki *et al.* (2014) |
|---|---|---|
|  | 30 GHz | |
| $\varepsilon'$ | 17.7 | 17.6 |
| $\varepsilon''$ | 16.5 | 15.4 |
| $L$ (mm) | 0.43 | 0.47 |
| $T_{tr}$ | 0.54 | 0.54 |
| 100 GHz |  |  |
| $\varepsilon'$ | 7.2 | 6.6 |
| $\varepsilon''$ | 8.3 | 8.0 |
| $L$ (mm) | 0.18 | 0.17 |
| $T_{tr}$ | 0.7 | 0.68 |

Several studies have directly measured the dielectric properties of skin above 6 GHz (Table 5). Most of these studies have used coaxial probes or open-ended waveguides placed against the skin or, at THz frequencies, measured wave reflection properties of the skin surface. These methods are sensitive to the dielectric properties of skin averaged over tissue volumes exposed to the field, typically including the stratum corneum, epidermis, and parts of the dermis. Some investigators have estimated properties of individual skin layers from such data using electromagnetic scattering models. Sasaki and his group (Sasaki *et al.*, 2014; Sasaki *et al.*, 2017) extended measurements up to 1 THz by using a combination of coaxial probe and wave reflection techniques. That group has also reported dielectric properties of rabbit and porcine ocular tissues up to 110 GHz (Sasaki *et al.*, 2015).

Gao *et al.* (2018) reported an extensive set of measurements of the dielectric properties of skin (26.5 to 40 GHz), using an open-ended waveguide placed against the skin of three human subjects in several locations (forearm, shoulder, abdomen, thigh, calf, palm). The investigators estimated the dielectric properties of different skin layers using an electromagnetic scattering model. In addition, they provided an extensive analysis of experimental uncertainties in measuring dielectric properties of skin in the mm-wave range. The study found that the permittivity of skin at 30 GHz varies by more than a factor of 3 in different sites of the body, reflecting variations in skin thickness; in addition, there was considerable intersubject variability (Table 5).

Despite these several studies, there remains limited data from skin and ocular tissues. The currently available data are sufficient for many electromagnetic modelling studies, but the data remain insufficient at mm-wave frequencies to explore the variations introduced by interpersonal and intrapersonal variations in RF absorption in skin and subcutaneous tissue.

Table 5. Selected studies on dielectric properties of tissues at frequencies > 6 GHz.

| References | Animal/Tissue (Frequency Range) | Comments |
|---|---|---|
| Sasaki *et al.* (2014) | Porcine epidermis and dermis *in vitro* at frequencies from 0.5 to 110 GHz. | Combination of dielectric probe (0.5 GHz−50 GHz) and free field techniques (50 GHz−110 GHz). |
| Sasaki *et al.* (2015) | Porcine and rabbit ocular tissues *in vitro* at frequencies from 0.5-110 GHz | Dielectric probe was used. Dielectric properties of the cornea, lens cortex, lens nucleus, aqueous humour, vitreous humour, sclera, and iris were reported. Rabbit tissues were typically used, but porcine tissue was used for the aqueous humour. The dielectric properties of vitreous and aqueous humour are almost equivalent to those of pure water at frequencies over around 10 GHz. |
| Sasaki *et al.* (2017) | Porcine dermis, subcutaneous tissue fat, muscle *in vitro* at frequency ranges of 100 GHz−1 THz, 1 GHz−1 THz, and 1−100 GHz, respectively. | Combination of dielectric probe (1 GHz−100 GHz) and free field techniques (100 GHz−1 THz). Study also included extensive Monte Carlo dosimetric modeling. |
| Gao *et al.* (2018) | Human skin *in vivo* at frequencies from 26.5 to 40 GHz. | Reflection measurements using open-ended waveguide probe. 3 human subjects in several locations: forearm, shoulder, abdomen, thigh, calf, and palm. Error analysis considering effects of thicknesses of skin layers. The reported relative permittivity of skin at 30 GHz (average of measurements on 3 subjects): forearm 17.1, shoulder 20.2, abdomen 20.5, thigh 16.7, calf 16.1, 2 sites on the palm 6.4, 9.2 GHz. |

*C. Bioheat Transfer Equation*

To assess thermal hazards from mm-wave exposure, RF-induced temperature increases in skin and cornea are of principal interest, though whole-body exposure will be mentioned briefly in the discussion section. Several studies have combined electromagnetic and thermal modeling to estimate the increase in tissue temperature due to mm-wave exposure. Nearly all thermal modeling studies have used the finite difference time domain method for electromagnetic modeling, with solution of Pennes' bioheat equation (BHTE) (Pennes, 1948) for thermal analysis. The BHTE equation can be written:

$$\nabla(k(\mathbf{r})\nabla T(\mathbf{r},t)) + \rho(\mathbf{r})SAR(\mathbf{r},t) + M(\mathbf{r},t) - \omega_b(\mathbf{r},t)\rho_b{}^2 c_b(T(\mathbf{r},t) - T_b) = c(\mathbf{r})\rho(\mathbf{r})\frac{\partial T(\mathbf{r},t)}{\partial t} \quad (1)$$

where $T$ is temperature of a control volume of tissue that is sufficiently small that the temperature throughout the volume can be considered to be uniform. Thermal properties for the tissue are considered to be averaged over this control volume and may be different from those of blood; $T_b$ is temperature of the perfusing blood (more precisely, the temperature of the arterial blood supply to a region of tissue, which is assumed to be constant and equal to the core body temperature); $c$ ($c_b$) is the specific heat of the tissue (blood) (J/kg/K); $\rho$ ($\rho_b$ is the density of the tissue (blood) (kg/m$^3$); $k$ is the thermal conductivity of the tissue (W/m/K); $M$ is the rate of metabolic heat generation (W); $\omega_b$ is the rate of blood perfusion per kg of tissue mass (m$^3$/(s kg)); $SAR$ is the specific absorption rate (SAR) (W/kg); and $t$ is time.

Heat exchange at the body surface proceeds through four main mechanisms: convective cooling by air, flowing past the body, evaporation of skin moisture, thermal conduction to surrounding materials, and radiation from the body surface back into space. The first of these processes is the dominant mechanism under ordinary conditions. It can be modeled through boundary condition at the skin-air interface:

$$-k\frac{dT_{sur}}{dz} = h(T_{skin} - T_{air}) \quad (2)$$

where $z$ is the axis normal to the surface and $h$ is a convective heat exchange coefficient that depends on air velocity and temperature. For a fuller description of the boundary conditions (a rather complex subject) (Xu *et al.*, 2009). For the presently considered frequency range, heat transport near the skin surface is dominated by thermal conduction into the tissue due to the high temperature gradients at the skin, and only a small fraction of the absorbed energy is lost back into the surrounding environment. Consequently, adiabatic boundary conditions ($h$=0 in Eq. 2) are usually a good approximation.

Eq. 1 is readily solved numerically as an extension to FDTD electromagnetic modeling, and thermal modeling programs are included in several major electromagnetic modeling programs (e.g. Semcad X, Schmid & Partner

Engineering AG, Zurich, and XFdtd, Remcom, State College PA). A variety of high-resolution image-based models of the human body are also available. Most thermal modeling studies discussed below computed the steady state increase in temperature (from Eq. 1 with the right-hand term in dT/dt set to zero) while a few studies computed the time-dependent increase in tissue temperature. For short times (a minute or less), effects of the blood perfusion term on the computed temperature increase are minor and Eq. 1 can be replaced by a simple heat conduction equation. Steady state is reached after several hundred seconds and the temperature increase is dominated by tissue blood perfusion (fourth term on the left side of Eq. 1) which is highly variable.

Extensive tables of thermal properties of tissues can be found at (Diller *et al.*, 2000; Duck, 1990; Hasgall *et al.*, 2015). However, these values are subject to considerable variability as discussed below.

In assessing thermal impacts of exposure to RF energy, a useful concept is heating factor, defined as the increase in temperature in the steady state per unit of exposure, with typical values of about 0.15 °C W$^{-1}$ kg below 3 GHz and 0.018 °C W$^{-1}$ m$^2$ at higher frequencies (Funahashi *et al.*, 2018b).

*D. Baseline model*

Foster and colleagues (Foster *et al.*, 2016; Foster *et al.*, 2018a; Foster *et al.*, 2018b; Ziskin *et al.*, 2018) have developed simplified "baseline" one-dimensional models for mm-wave heating of tissue which provides insights into the thermal response. A simplified version of Eq. (1) can be written in form

$$k\nabla^2(\delta T) - \omega\rho^2 c(\delta T) + \rho SAR = \rho c \frac{d(\delta T)}{dt} \qquad (3)$$

where δ*T* is the temperature increase above the baseline (pre-exposure) value.

The model consists of an insulated half-space of material with thermal and electrical properties characteristic of skin, exposed to normally incident plane wave energy. For a plane wave of intensity *I*$_o$(*t*) incident on a planar surface, the absorbed power density (SAR) at a depth *z* beneath the surface is

$$SAR = \frac{I_o T_{tr}}{\rho L} e^{-z/L} \qquad (4)$$

where $L$ is the energy penetration depth in tissue and $T_{tr}$ is the energy transmission coefficient into tissue. Eq. 3 has two intrinsic time scales:

$$\tau_1 = \frac{1}{\omega_b \rho} \quad (5a)$$

$$\tau_2 = \frac{L^2 \rho c}{k} \quad (5b)$$

The first of these ($\tau_1$) characterizes the rate of removal of heat from subcutaneous tissues to the central core of the body by blood perfusion, and is ≈500 sec for typically reported values of skin blood flow. The second ($\tau_2$, < 2 sec at mm-wave frequencies) represents the rate of thermal diffusion from the exposed layer of tissue into deeper tissues. The increase in surface temperature $\delta T_{sur}$ is determined by the interplay between the rate of heat generation in the skin layer exposed to radiofrequency radiation, the rate of diffusion of heat from the exposed layer into deeper tissues (a relatively fast process), and the rate of removal of heat from deeper tissues to the body core by blood perfusion (a much slower process). Skin is a very leaky reservoir for heat due to its small thickness, but if energy is pushed into it sufficiently rapidly (i.e. if the incident power density is high), significant temperatures increases can develop. For more modest heating rates, the increase in skin temperature will be determined instead by thermal washout in tissues beneath the skin layer.

The simple baseline model (Eq. 3) yields analytical solutions for simple cases, but in general these are unwieldy. Foster et al. have developed approximate solutions for the increase in surface temperature that apply at mm-wave frequencies.

Surface heating approximation ($L \to 0$). The increase in surface temperature $\delta T_{sur}$ can be written:

$$\delta T_{sur} = \frac{I_o T_{tr} L}{k} \sqrt{\frac{\tau_1}{\tau_2}} \, erf\left(\sqrt{\frac{t}{\tau_2}}\right) \quad (6)$$

which approaches the steady state temperature

$$\delta T_{sur} \to \frac{I_o T_{tr}}{\rho \sqrt{k \omega_b c}} \quad as \ t \to \infty \quad (7)$$

For $\tau_2 \ll \tau_1$ this is very close to the steady state increase in surface temperature provided by the full analytical solution to Eq. 3 for the adiabatic plane. In this approximation, the increase in surface temperature in the steady state scales as $(\omega_b)^{-1/2}$. However, this model behaves poorly for calculating early transient temperature increases from intense pulses (in mathematical terms, its impulse response diverges). Physically, this is a consequence of pushing a finite amount of energy into an infinitesimally thin tissue layer., However this approximation works quite well for exposure times more than a few seconds for mm-waves (Foster *et al.*, 2020).

Conduction only model ($\omega_b = 0$), which applies at short times after exposure has begun where heat conduction in the skin layer is the dominant mode of heat transfer:

$$\delta T_{sur}(t) = \frac{I_o T_{tr} L}{k} \left[ 2\sqrt{\frac{t}{\pi \tau_2}} + \left( e^{\frac{t}{\tau_2}} erfc\left(\sqrt{t/\tau_2}\right) - 1 \right) \right]$$

$$\approx I_o T_{tr} \left( 2\sqrt{\frac{t}{\pi \rho c k}} - \frac{L}{k} \right) \quad (t > \tau_2) \tag{8}$$

$$\to \frac{I_o T_{tr} t}{k L \rho c} \text{ as } t/\tau_2 \to 0$$

Eq. 8 is an excellent approximation of the full analytical solution to the model for short times after exposure is begun, before thermal convection due to blood perfusion becomes significant. Also, its asymptotic approximations are also presented. Because it neglects effects of blood perfusion, which becomes the main mode of heat transfer as time progresses, Eq. 8 does not have a finite steady-state solution in the 1D model using a semi-infinite plane of tissue. However, any realistic finite-sized model would impose boundary conditions that would result in a finite steady-state temperature increase.

These simplified solutions suggest that for exposure times much shorter than the thermal time constant τ2 in (5b) the increase is particularly sensitive to the product ρck (the thermal inertia of the tissue) (Eq. 8). Close to the steady-state, the temperature increase at the surface scales as $(\omega_b)^{-1/2}$ (Eq. 7). These results apply to the simplified 1D "baseline" model but give an approximate picture of the behavior of more detailed models as well.

## 3. Physical Quantities for Local Exposure Below and Above 6 GHz

*A. Guidelines and Standard*

In the guidelines/standard, first, a physical quantity related to exposure to RF energy (including averaging region) is derived to correlate with the temperature rise, and then the corresponding limit/restriction is derived. At frequencies below 3-10 GHz, the SAR averaged over 10 g of tissue is an approximate surrogate of local temperature rise (Hirata *et al.*, 2006; Razmadze *et al.*, 2009; McIntosh and Anderson, 2010b) (see also the review by Foster *et al.* (2018a)). The revised IEEE and ICNIRP limits introduced a more useful surrogate for local temperature rise for use at frequencies > 6 GHz, the absorbed/epithelial power density (i.e. the power density absorbed in tissue from an incident RF wave, in terms of watts of power per unit area of skin). This power density is to be averaged over a specific area (averaging area) (International Commission on Non-Ionizing Radiation Protection, 2020; IEEE-C95.1, 2019).

To derive an appropriate averaging area, we note that the side length of the cube corresponding to 10 g tissue (the averaging volume below 6 GHz) is 2.2 cm, corresponding to a square with area 4.8 cm$^2$ (assuming that the density of the tissue is 1,000 kg/m$^3$. To interface smoothly with the limits below 6 GHz, this suggests a choice of averaging area of about that size. IEEE C95.1-2019 specifies that "*the choice of 4 cm$^2$ (for ERL between 6 GHz and 300 GHz) and 1 cm$^2$ (for high-power pulsed RF exposures) for the spatial peak averaging area was influenced by several factors. First, there is general agreement with other guidelines and standards including ICNIRP and ANSI Z136.1-2014. Second, the smaller averaging area (1 cm$^2$) provides an additional level of conservatism for brief, high-power pulses…*". The ICNIRP guidelines specify as follows: "*… ICNIRP uses a square averaging area of 4 cm$^2$ for >6 to 300 GHz as a practical protection specification. Moreover, from >30 to 300 GHz (where focal beam exposure can occur), an additional spatial average of 1 cm$^2$ is used to ensure that the operational adverse health effect thresholds are not exceeded over smaller regions…*" for a limit relaxed by a factor of 2.

*B. Review of Studies on Averaging Area of Absorbed Power Density*

This section reviews studies on the relationship between the power-density averaging area and the peak increase in tissue temperature for exposures above 6 GHz, including the rationale of averaging area. Foster *et al.* (2017) suggested a distance where the temperature increase is removed by blood flow is 7 mm, approximately corresponding to a circle with a diameter of 14 mm. This distance is derived for a homogeneous skin model whose thermal parameters are in Hasgall *et al.* (2015). . This area is smaller than the 10 cm$^2$ (frequency independent) used in previous international guidelines and standards (ICNIRP, 1998). The averaging area of approximately 4 cm$^2$ was suggested from FDTD computations (Hashimoto *et al.*, 2017). For beam exposure smaller than 4 cm$^2$, those investigators proposed a compensation scheme to estimate the skin temperature, based on Foster *et al.* (2017). He *et al.* (2018), using FDTD analysis for realistic antennas for the 5$^{th}$ generation wireless communications, proposed a similar averaging area.

Neufeld *et al.* (2018) derived an averaging area to limit the temperature rise to 1 ºC for a spatially nonuniform incident power density whose averaged intensity was 10 W/m$^2$. Those authors defined the averaging area as functions of frequency and distance from the transmitter to the body. Unlike other studies, Neufeld *et al.* (2018) considered near-field exposures, (including cases where the distance from the transmitter to skin surface was larger than 2 mm.

In sum, the latest versions of both IEEE and ICNIRP limits specify that the absorbed power density must be averaged over tissue areas of 4 cm$^2$ above 6 GHz. At mm-wave frequencies, where beams smaller than 4 cm$^2$ in area may be feasible, a smaller averaging area might be better correlated with the spatial-peak temperature rise. However, extreme exposure situations such as considered by Neufeld *et al.* (2018) may require different choices of averaging area. Those are most likely to occur when a small antenna operating at mm-wave frequencies is in close proximity to the skin.

## 4. Steady-State Temperature Rise in Skin above 6 GHz

### A. *Review of Computational Dosimetry*

Studies reporting on computational studies and specific metrics for temperature rise (e.g. incident power density) were included. In total, 14 studies were reviewed. The criteria for paper selection can be found in the Appendix (to be listed; currently only for search strategy).

Gustrau and Bahr (2002) measured the dielectric properties of eye and skin tissues between 75 and 100 GHz and conducted electromagnetic and thermal simulations in for skin and a detailed model of the human eye subject to exposure to plane waves at 77 GHz. Thermal measurements *in vivo* in skin (in the forearms of two volunteers) and *in vitro* in excised porcine eyes, respectively, showed steady-state temperature increases of 0.7 ºC from exposure at an incident power density of 100 Wm$^{-2}$, which were consistent with simulation results "in view of the natural variability of the measurement data… and reduced complexity of the model" (Gustrau and Bahr (2002)).

Kanezaki *et al.* (2009) derived an approximate expression for thermal steady state temperature rise in the skin layer of a three-layer (skin, fat, and muscle) one-dimensional model exposed to a plane wave at frequencies from 30 to 300 GHz. A Debye-type approximation was introduced to model the dielectric properties of the tissues between 30–300 GHz. The authors concluded that the effects of variations in dielectric properties on skin heating were smaller than from variations in the assumed thermal parameters and thickness of the tissue layers. Skin heating correlated with the power density absorbed in the skin rather than the SAR at the skin surface or the incident power density on the skin.

The same group derived the temperature rise in the thermal steady state in the skin layer of a one-dimensional one-layer (skin) and three-layer (skin, fat, and muscle) model for plane wave exposure at frequencies 30–300 GHz (Kanezaki *et al.*, 2010). The peak temperature was 1.1 ºC at 30 GHz and 1.9 ºC at 300 GHz with an incident power density of 50 W m$^{-2}$. The surface temperature rises in the three-layer model were 1.3–2.8 times greater than those in the one-layer model due to the thermal resistance of the fat layer. The heat transfer coefficient $h$ (cf. Eq. 2) was the most dominant parameter in the change of the surface temperature rise in the three-layer model.

Table I. Computational studies on temperature rise: Steady-state exposures

| References | Study Design Frequency, exposure | Major Findings/comments |
|---|---|---|
| Gustrau and Bahr (2002) | Analytic model for SAR in skin, 3-100 GHz; FDTD calculation of SAR in skin and eye at 77 GHz in vitro (eye) and in vivo (forearm of 2 subjects). | The results of the thermal measurements and simulations provide consistent results for the assessment of thermal effects. Used literature values for skin blood perfusion, experimental temperature increase in rough agreement with model. |
| Kanezaki *et al.* (2009) and Kanezaki *et al.* (2010) | Analytical/numerical model (SAR in skin and skin layers and temperature increase in skin and eye using BHTE), 30-300 GHz. | The incident power density of 50 W/m$^2$ causes temperature increase of 0.6–0.9 °C. The fat causes higher computed steady-state temperature increases due to the adiabatic nature. |
| Morimoto *et al.* (2016) | FDTD simulation, 1-30 GHz, dipole antenna | The computed steady-state temperature rises in the head (skin) increase with the increase of the frequency. The SAR averaged over 10 g of tissue depend on the averaging schemes. |
| Laakso *et al.* (2017b) | FDTD calculation in anatomical human head models, 1-12 GHz, independently determined tissue blood perfusion | Variability of peak computed steady-state temperature rise in the skin due to individual and regional variations in the blood flow was less than ±15%. |
| Leduc and Zhadobov (2017) | 60 GHz phantom in vitro and 1D model of surface heating | 50 W/m$^2$ causes computed steady-state temperature increase of 0.6-0.9 °C. Comparison to analytical solutions of heat conduction equation. |
| Sasaki *et al.* (2017) | Simulation study, skin heating (steady state) 10 GHz – 1 THz, Detailed Monte Carlo model | For variation of human tissue composition, the computed steady-state temperature rise at the skin surface for different tissue thickness is generally 0.02 °C or less for an incident power density of 1 W/m$^2$ at a frequency below 300 GHz. |
| Funahashi *et al.* (2018b) and Funahashi *et al.* (2018c) | FDTD simulation, 0.3-300 GHz, dipole, patch antennas | The absorbed power density is a good metric of computed steady-state temperature rise in the skin from 30–300 GHz, and provides a less accurate but still conservative estimate down to 10 GHz, whereas the SAR is a good metric below 3 GHz. The heating factor for plane wave exposure is 0.15 °C W$^{-1}$ kg below 3 GHz and 0.018 °C W$^{-1}$ m$^2$ above 10 GHz. |

| Kodera *et al.* (2018a) | FDTD simulation 300 MHz–10 GHz, head and brain heating, includes vasodilation in model. | The effect of vasodilation is significant, especially at higher frequencies where the highest increase in tissue temperature occurs in the skin. Its effects become notable at an SAR >10 W/kg. |
|---|---|---|
| Ziskin *et al.* (2018) | Simulation study at 6–100 GHz, skin heating. BHTE incorporating blood flow rate-dependent thermal conductivity | The computed steady-state temperature increase at the skin surface is determined by the thermal resistance of subcutaneous tissues, blood flow in the dermal and muscle layers, and thickness of subcutaneous fat. |
| Li *et al.* (2019b) | Human skin exposure to obliquely incident electromagnetic waves at frequencies from 6 GHz to 1 THz. Monte Carlo analysis | The absorbed power density provides an excellent estimate of computed steady-state skin temperature elevation through the millimeter-wave band (30-300 GHz) and a reasonable and conservative estimate down to 10 GHz, whereas the SAR is a good metric below 3 GHz. |
| Nakae *et al.* (2020) | FDTD simulation at 28 GHz. Cubic model, 4 and 8 element dipole arrays. | The enhancement of the ratio of the computed steady-state temperature increase to incident power density was observed around the Brewster's angle. |
| Christ *et al.* (2020) | Simulation study, skin heating (steady state) 6 GHz – 100 GHz | When the stratum corneum and related layers serve as matching layer that increases the power absorption and the resulting computed temperature increases in the tissue. |

Morimoto *et al.* (2016) computed the SAR and steady state temperature rise in the head and brain from 1 GHz to 30 GHz. As the frequency increases, the computed temperature rise in the head increase, and in the brain decrease, due to absorption of energy progressively closer to the body surface. Morimoto et al (2016) noted that SAR averaging algorithms excluding the pinna must be used when correlating the peak temperature elevation in the head excluding the pinna.

Sasaki *et al.* (2017) measured the dielectric properties of tissue and computed the steady-state temperature rise in a one-dimensional four-layer model (epidermis, dermis, subcutaneous tissue, and muscle) exposed to plane wave RF energy at 10 GHz– 1 THz. Using a Monte-Carlo simulation, they studied the variations of temperature rise due to variations in thickness of the tissue layers obtained from statistical data from the human body. The computed steady state temperature rise over the skin surface was generally < 0.02 °C for an incident power density of 1 W m$^{-2}$ at frequency below 300 GHz. From the models, the investigators determined the maximum incident power density vs. frequency that would result in steady-state temperature rises within specified limits.

Ziskin *et al.* (2018) developed a simplified model for skin that incorporates anatomic detail, using a series of planar structures representing the skin and subcutaneous fat. The model shows that the thermal resistance of subcutaneous fat contributes significantly to the steady-state increase in computed skin temperature.

Funahashi *et al.* (2018a) studied both analytically and computationally the effectiveness of the absorbed power density at the skin as a metric to estimate the steady-state rise in skin temperature at frequencies above 6 GHz. They concluded that the absorbed power density provided an excellent estimate of skin temperature rise through the millimetre-wave band (30–300 GHz) and provided a less accurate but conservative estimate down to 10 GHz, whereas the SAR is a good metric below 3 GHz. They reported that the heating factor for plane wave exposure is 0. 15 °C $W^{-1}$ kg below 3 GHz and 0.018 ºC $W^{-1}$ $m^2$ above 10 GHz. Funahashi *et al.* (2018c) confirmed the effectiveness of the absorbed power density for use in estimating the skin temperature rise even for a realistic antenna but noted that for small beam diameters an averaging area smaller than 4 $cm^2$ is needed above 30 GHz.

Li *et al.* (2019a) analyzed the temperature rise in the thermal steady state in skin exposed to plane waves with oblique incidence at frequencies from 6 GHz–1 THz, using the four-layer plane model described in (Sasaki *et al.*, 2017). The investigators studied the variations in computed temperature rise and total power transmittance into the skin as functions of angle of incidence and wave polarization. For incident waves with transverse magnetic (TM) polarization, the transmittance increased with the angle of incidence because of the Brewster effect. The computed temperature rise produced by waves with oblique incidence never exceeded that of normally incident waves of the same power density.

Nakae *et al.* (2020) computed the temperature rise in the thermal steady state for dipole arrays at 28 GHz. The investigators reported an increased ratio of the temperature increase to incident power density for incidence angles of radiation agrees with Li *et al.* (2019b), but for only the angle range close to the Brewster's angle appear disagreement. For the dipole antenna arrays, is the distances between the antenna and body that were large (e.g., 45 mm) for large incident angle, which is not realistic for compliance assessment of transmitters operated near the body. This study presented for a given output power, the highest absorption, consistent with Li *et al.* (2019b), is when the beams impact the tissue with normal incidence.

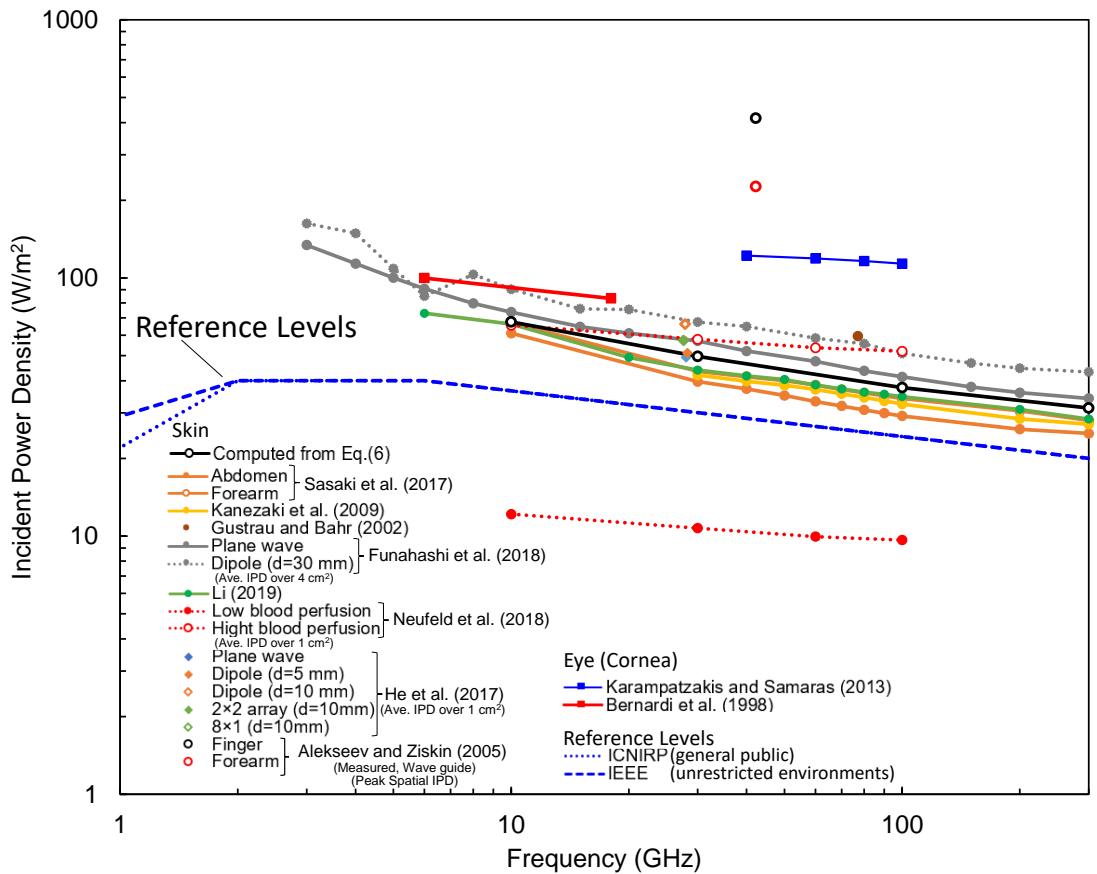

Fig. 1. Incident power density needed to increase the skin/cornea temperature by 0.5 °C in the steady state, as computed by thermal modeling programs.

Christ *et al.* (2020) computed the steady-state temperature rise in multilayer skin models including the stratum corneum and the viable epidermis, over a frequency range from 6 to 100 GHz. Under "worst case" assumptions, i.e. that the thickness and dielectric properties of the tissue layers adjusted to maximize transmission into the skin and adiabatic boundary conditions at the air-surface boundary, the authors reported that the calculated steady-state temperature rise at the surface is 0.4°C for an incident power density of 10 W/m$^2$. For other exposure scenarios, the same incident power density produced calculated temperature increases of 0.1–0.2°C (for a thin stratum corneum) and 0.1–0.3 °C for a thick stratum corneum.

*B. Power Density and Temperature Rise*

Figure 1 summarizes the required power densities from 1 to 300 GHz reported in several studies to increase skin/cornea temperature in the steady state by 0.5 °C; virtually all of these studies found that exposure levels exceeding occupational limits of IEEE and ICNIRP would be required. Only one study (Neufeld *et al.*, 2018) reported that incident power densities below international limits could cause a 0.5 °C increase in the steady state. That result was from a worst-case scenario, in which the thickness and dielectric properties of cutaneous and subcutaneous tissue layers had been chosen to maximize the fraction of absorbed power, and is unlikely to represent a realistic exposure situation.

Equation 6 predicts that an incident wave of 100 $W/m^2$ will produce steady-state temperature increases ranging from 0.74 °C (10 GHz) to 1.58 °C (300 GHz) in a uniform 1D baseline model. More complex models produce similar results. For a multilayer model for the forearm and abdomen, (Sasaki *et al.*, 2017) calculated temperature increases (scaled to an incident power density of 100 $W/m^2$) of 0.74 and 1.76 °C, respectively, for the forearm at 10 and 300 GHz, and 0.84 and 1.90 °C for the abdomen.

All of the results discussed above are from computational studies. Only one study in Table reported experimental measurements of temperature increases in human subjects from exposures to RF energy at frequencies above 6 GHz. Alekseev *et al.* (2005) measured the rise in skin temperature exposed to RF energy at 42.25 GHz, from a rectangular waveguide antenna. In that study, the antenna produced a circularly symmetric Gaussian exposure pattern on the skin. A peak incident power density was 2080 $W/m^2$ on finger and forearm of human subjects. The IPD averaged over 4$cm^2$ estimated from the IPD distribution on the skin surface was 347.6 $W/m^2$. The value plotted in Fig. 1 was estimated as 69.5 $W/m^2$ in finger and 37.8 $W/m^2$ in forearm, respectively, assuming the Gaussian field distribution. Based on the theory in Foster *et al.* (2016), Hashimoto *et al.* (2017) suggested an equivalent power density for a smaller beam in terms of the root of the ratio of the exposure area to 4 $cm^2$. The corresponding incident power density over 4 $cm^2$ would be 795.6 $W/m^2$ with a factor of 0.44. The plots in Fig. 1 would be lowered by a factor of 0.44 from the original value; 159 $W/m^2$ in finger and 86 $W/m^2$ in forearm, respectively, which are in good agreement with other computational analysis.

Other studies have been reported by Japanese groups funded by Ministry of Internal Affairs and Communications, Japan. Kodera *et al.* (2017) conducted FDTD analysis for rat head (brain) exposed to RF radiation at 6 and 10 GHz from antennas near the body surface. The experimentally measured and computed temperature rises were in good agreement, considering the effects of vasodilatation predicted by a thermoregulatory model. Straightforward comparison of these results with other studies is not feasible, because the SAR pattern in the brains of the rats was inhomogeneous; the study used the brain - averaged SAR as the metric for evaluation which also prevents comparison with studies on larger animals.

Kojima *et al.* (2018) exposed the eyes of rabbits to millimeter waves at 40, 75, and 95 GHz (extended in 2020 to 162 GHz) and documented damage to the eyes (as well as eyelids) depending on exposure (Kojima *et al.*, 2020). Those authors measured increases in corneal temperature during the exposure; the initial temperature was 34.5 °C and then reached 37.6 °C at 6 min for exposure with 500 W/m$^2$ at 75 GHz. So far, no detailed thermal modeling of this exposure situation has been reported.

TABLE II *Computational Studies Including Temperature Increase: Brief Exposures*

| Study | Study Design | Major Findings |
|---|---|---|
| Foster *et al.* (1998) | Analytical modeling study based on simplified BHTE | Experimental agreement with estimated threshold for perception or pain for plane-wave irradiation as a function of frequency and exposure duration. One-dimensional model provides conservative estimation for extreme heating situation (exposure to brief high fluence pulses). |
| Nelson *et al.* (2000) | Modeling study only (layered spherical model for head of monkey) 100 GHz, up to 30 kW/m$^2$ for 3 sec or 3kW/m$^2$ for 30 sec. Studied effects on: 1) surface convection coefficient; 2) surface evaporation rate (i.e., sweating); 3) blood-flow rate to the scalp/surface tissue | The peak surface temperature is affected by environmental conditions (convection coefficient, sweat rate). Subsurface temperature increases are considerably lower than increases in surface temperatures. |
| Alekseev *et al.* (2005) | Experimental study, forearm and middle finger skin exposed from open ended waveguide at 42.25 GHz. Two-dimensional computational modeling | Local heating of the skin was greatly reduced by elevated blood perfusion occurring in the forearm and in the finger. The relationship between blood |

| | based on BHTE with effective thermal conductivity. | flow and the effective thermal conductivity was linear. |
|---|---|---|
| Morimoto et al. (2017) | Modeling (FDTD) 1-30 GHz, layered plane and head models, effects of beam diameter, up to 2000 sec | Calculated temperature elevation at the skin surface for short pulse exposure (<10 s, beam of 20 mm) is at least twice higher (15–30 GHz) compared to that produced by continuous exposure. Shorter thermal time constant with higher radiation frequency (16 and 5 min at 1 and 30 GHz). |
| Laakso et al. (2017a) | Modeling (FDTD), 6-100 GHz, pulses < 10 sec. Human face model. Considered maximum-fluence pulses consistent with earlier IEEE and ICNIRP limits | Areas of enhanced absorption near edges of eye and nose, due to complexity of the surface (<10 s). Effect of pulsed exposure duration diminishes as the frequency decreases. |
| Foster et al. (2018b) | Analytical modeling study based on simplified BHTE | The impulse response to millimeter wave radiation (30-300 GHz) showed a sharp peak temperature rise due to short term accumulation of heat near the surface. |
| Kodera et al. (2018b) | Modeling (FDTD) 0.1-6 GHz, layered plane and human head models, exposure from 0.01 sec to 6 min and pulse train at frequencies 0.1–6 GHz | Maximum temperature rise (brief intense exposure with a total fluence corresponding to the maximum allowable SAR averaged over a 6-min averaging time) exceeded the steady-state temperature above 400 MHz (continuous exposure with the same time-averaged SAR) |
| Neufeld and Kuster (2018) | Analytical approach applied to peak temperature increase in the skin for plane-wave and localized exposures (< 600 min) | Estimation of a maximum averaging time of 240 s for mm-waves based on surface heating theory to limit the maximum local temperature increase to 1 °C for pulses of duty cycle ≥ 0.1. |

## 5. Temperature Rise After Brief Exposures

A. Review of Computational Dosimetry

Studies that have investigated the effects of brief pulses or sequences of brief pulses incident on skin have are reviewed in this section. For exposure times much shorter than the thermal time constant $\tau_2$ in (5b), the surface temperature increases almost linearly with time in accordance with Eq. (8). Nine papers were included in the review.

Foster *et al.* (1998) proposed an equation for an upper-limit increase in skin temperature assuming a one-dimensional model that applies in extreme heating situations, e.g. if all of the exposure during the averaging time occurs in one brief pulse, and estimated thresholds for perception or pain for plane-wave irradiation as a function of frequency (1-300 GHz) and exposure duration. They also discussed how the microwave and laser standards differ in their formulation, particularly with respect to thermal averaging time. (This early study has largely been supplanted by more recent and more detailed studies by Foster et al. cited elsewhere in this review).

Nelson *et al.* (2000) calculated the temperature rise in a spherical four-layer model of the body produced by exposure to intense RF pulses (100 GHz) of duration 3 s (10-30 kW m$^{-2}$) and 30 s (1-3 kW m$^{-2}$). In both cases, the applied energy densities (pulse fluences) were 30-90 kJ/cm$^2$. The calculated increase in skin temperature was 22–24ºC from a 3 s pulse and 7–12ºC from a 30 s pulse, in each case with a fluence of 90 kJ m$^{-2}$.

Morimoto *et al.* (2017) computed the thermal time constant, defined in that study as the time for the peak temperature increase to fall by a factor of 1/e (approximately $\tau_1$ in (5a)), in anatomically detailed image-based models of the human body for exposure frequencies up to 30 GHz. They showed that the thermal time constant declines with increase in frequency to reach 16 min at 1 GHz and 5 min at 30 GHz, respectively. These changes result from different depths of energy penetration. Deep tissues (e.g. brain) have a slower thermal response than superficial tissues due to the fact that most of the RF energy is deposited outside of the skull in the scalp and the deposited heat has to be conducted through the skull into brain tissue, which increases the thermal response time of the brain to 10–30 min.

Laakso *et al.* (2017a) computed the temperature rise in the human face produced by plane wave pulses shorter than 10 s at frequencies from 6 to 100 GHz. The time constants that characterize the rate of temperature rise depended on the three-dimensional distribution of energy absorption, with more localized absorption near the body surface resulting in a more rapid initial increase in temperature and shorter thermal response times. Laakso et al. (2017a) showed that the peak temperature rise was below 1.5ºC for a pulses of < 0.1 s duration and fluence of 1 kJm$^{-2}$.

Foster *et al.* (2018b) compared the temperature rise produced by a single RF pulse or pulse train with a pulse duration of 0.57 ms to 1000 s at frequencies 1–300 GHz from predictions of the 1D baseline models and more detailed calculations based on an anatomical human head model. The radiofrequency radiation consisted of both plane waves incident on the head, and radiation from a dipole antenna close to the head. The impulse response to millimetre wave radiation (30-300 GHz) showed a sharp peak temperature rise due to short term accumulation of heat near the surface.

Kodera *et al.* (2018b) investigated the transient temperature rise from exposure to pulses of 0.01 sec to 6 min, considering individual pulses and pulse trains, at frequencies 0.1–6 GHz. The maximum transient temperature rise produced by a maximally intense pulse (with the maximum fluence permitted by the exposure limit times the 6-min averaging time) exceeded the steady state temperature rise produced by continuous exposure with the same time-averaged SAR, at frequencies above 400 MHz. These authors subsequently extended their studies to the frequency range from 6-300 GHz in Kodera and Hirata (2019). Similar results have been reported by Foster et al (2019,2020).

Neufeld and Kuster (2018) carried out a worst-case thermal analysis of thermal response of tissue using the surface heating approximation for pulse-train exposures. They suggested a maximum averaging time of 240 s for mm-waves to limit the maximum local temperature increase to 1 °C for pulses of duty cycle $\geq 0.1$. The validity of the surface heating approximation for simulations of transient heating by brief intense pulses has been discussed Foster (2019) and Neufeld and Kuster (2019) (see Sec 4. B).

Foster *et al.* (2020) calculated transient temperature increases in the backs of human subjects exposed to high-fluence 3-sec RF pulses at 94 GHz. The measured temperature increases agreed well with predictions based on the 1D baseline model. Similar modeling of the transient temperature rise in the corneas of rabbits exposed to similar RF pulses (Foster *et al.*, 2003) was less successful due to the high variability in the temperature rises in the cornea, which is attributable to the presence of standing waves caused by scattering of RF energy from the eyelid.

B. Comparison of Computation and Analytic Solution

Figure 2 compares the transient temperature rise in skin for RF pulses with constant fluence of 36 kJ/m² for different pulse widths. The analytical solution of the 1D baseline model agree well numerical solution (FDTD method) for the model. However, the surface heating approximation overestimates the transient temperature increase for short pulses of duration < 1 s) due to the singularity in its transient response as discussed above. (The surface heating approximation is much more accurate for pulse widths > 1 s and particularly for the steady-state temperature increase). For short pulses the errors introduced by the surface heating approximation can be very large, e.g., 182 times for 10 $\mu$s, 18 times for 1 ms. Thus, while the surface heating approximation results in simple analytic results for the increase in surface temperature at mm-wave frequencies, its limitations for short pulses should be considered as commented in Foster (2019).

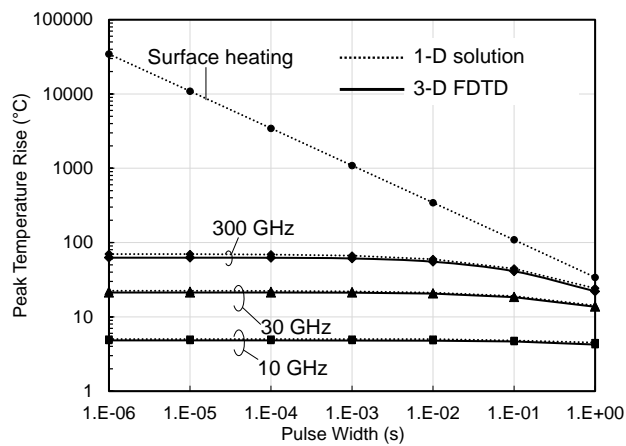

Fig. 2. Instantaneous temperature rises in a 1D model for skin, comparing analytical results for the model, the surface heating approximation, and a detailed numerical solution using the 3-D FDTD method. This figure shows the large errors introduced for short pulses (< 1 s) using the surface heating approximation.

C. Comparison of Computation and Measurement

Perhaps the most extensive measurements of the temperature increase from pulsed mm-waves were done under support of the U.S. Department of Defense beginning in the late 1990s as part of a program to develop an (as yet unused) nonlethal weapons system, called the Active Denial system (ADS) (Zohuri, 2019). The weapon beams brief (about 3 s duration) pulses of high intensity mm-waves (94 GHz) at targets with the aim of eliciting cutaneous thermal pain without causing thermal burns.

Figure 3 shows the measured temperature increase in rhesus cornea (Chalfin *et al.*, 2002; Parker *et al.*, 2020) and skin from human subjects (Walters *et al.*, 2000; Parker *et al.*, 2016) exposed to 2-4 sec. pulses of 94 GHz RF energy vs. calculated results from the 1D model. The calculations were done using the 1D baseline model with no adjustable parameters: the thermal parameters for dry skin and cornea were taken from the IT'IS dataset (Hasgall *et al.*, 2015) and the electrical parameters were calculated from dielectric data in (Gabriel *et al.*, 1996). Under the exposure conditions (short pulses) of these experiments, the conduction-only approximation (Eq. 7) closely agrees with the full solution to the bioheat equation.

Three of the sets of results (Walters *et al.*, 2000; Chalfin *et al.*, 2002) in Fig. 3 agree well with the predictions of the simple 1D model (Eq. 3), while the other (Parker *et al.*, 2016) diverges from the model. A larger scatter in the earlier set of data from the rhesus cornea results (Chalfin *et al.*, 2002) because that study recorded temperature increases across the whole cornea, which were strongly affected by standing wave effects, while the more recent study (Parker *et al.*, 2020) analyzed temperatures only from in the central 1/3 of the cornea to avoid interference effects (Foster et al. in press). The outlying data points for skin (Parker *et al.*, 2016) have no apparent explanation and dosimetry errors cannot be excluded.

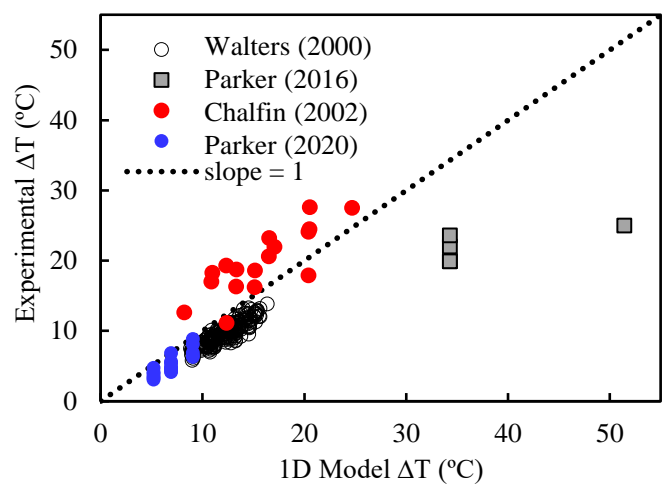

Fig. 3. Calculated vs. measured transient temperature increase in rhesus cornea (blue, red dots) and skin of human subjects (black circles, squares) from brief (2-4 s) pulses of 94 GHz radiation. Each point represents a single measurement. Calculations for the 1D model used literature values for dielectric and thermal parameters for skin and cornea without further adjustement.

## 6. Discussion

The following comments address remaining uncertainties in the modeling that should be addressed by future studies.

A. Parameter uncertainty

*Uncertainties in dielectric properties.* The dielectric properties of tissue determine the absorption of RF energy as well as reflection of energy from a tissue surface. There are, however, a number of important sources of uncertainty in these parameters. Variabilities in tissue composition (in particular tissue inhomogeneity and water content) contributes the variability in the tissue dielectric properties. In particular, the variabilities in tissue water content tends to be large in the tissues with low water content, and that to fat contributes by a factor of 3 of the tissue dielectric properties (Gabriel and Peyman, 2006; Sasaki *et al.*, 2017; Pollacco *et al.*, 2018). In addition, the variation of tissue dielectric constant with temperature in the range 6 – 100 GHz is marked, because of the strong effects of temperature on dielectric dispersion of water (the main constituent of tissue) in this range (Andryieuski et al., 2015). Most thermal modeling studies use parameter values for the dielectric properties of tissues taken from a few sources, which contributes to the consistency of results across studies but not necessarily to generalizability of results. These data were typically measured in excised animal tissues that may not accurately reflect the dielectric properties of human tissue *in vivo*. Moreover, most studies used a constant set of dielectric properties for skin. However skin is a heterogeneous tissue, and the water content of stratum corneum differs considerably from that of epidermis, and the relative thickness of the stratum corneum is highly variable across the body (Gao *et al.* (2018)). Consequently, the use of a lumped parameter to represent dielectric properties of "skin" will introduce considerable uncertainty. The variation of tissue dielectric properties by animal species and physiological condition needs further clarification.

*Uncertainties in material thermal properties* The material thermal properties of tissues (thermal conductivity, heat capacity, thermal diffusivity) used in modelling studies are mainly based on a few sources (Duck, 1990; Mcintosh and Anderson, 2010a; Hasgall et al., 2015). These properties were mostly measured in excised animal tissues. The thermal inertia of skin ($\rho k C_p$) can vary by 50% or more depending on the values for the individual parameters that are chosen in the literature (Hasgall et al., 2015). In addition, the thermal inertia of skin varies with location on the body and in different subjects, and increases by more than a factor of 4 with vasodilation (Lipkin and Hardy, 1954). Although skin temperature for most of the body is maintained within a narrow range, at the extremities the variation is much greater (Hardy *et al.*, 1938; Klinke *et al.*, 2009). The effects of ambient temperature on the tissue perfusion parameter $\omega_b$ in Equation 1 is usually linked to skin temperature (for example see (Moore et al., 2015; Fiala et al., 2001)), but the modelling of this parameter and the convective heat parameter *h* in Equation 2 may not fully capture extremes of ambient temperature. These variations will affect computed thermal response of tissues to RF radiation.

*Physiological variability* Reflecting its thermoregulatory function, skin is well supplied with blood vessels, and skin blood flow can vary by more than an order of magnitude depending on thermoregulatory status of an individual (ILO, 2020). Under ordinary room conditions, skin blood flow in humans varies by factors of 2–4 or more depending on the part of the body and the measurement technique (Hertzman and Randall, 1948). Skin blood flow in the head under resting conditions may vary by a factor of 3 and also depends on the direction relative to the skin surface because of the presence of the capillary bed (Laakso et al., 2017b). Since the steady state temperature scales as $\omega_b^{-1/2}$ this could lead to uncertainties of a factor of two or more in calculated temperature rises in skin, particularly in the steady state.

Anatomical variability Variability in tissue segmentation is another comparatively unexplored source of variability in thermal modeling studies. Sasaki *et al.* (2017) reported that variations in thickness of subcutaneous tissue layers among different individuals contributed more to variability in calculated steady-state temperature increases than measurement uncertainties of dielectric properties. Subcutaneous fat acts as a layer of thermal insulation and greatly affects the steady state increase in skin temperature (Ziskin *et al.*, 2018), and its thickness varies greatly among individuals and in different parts of the body. Individual features of the face or other body parts affect the distribution of the absorbed RF energy due to reflection and interference effects (Laakso *et al.*, 2017a). Their effects on inter-individual variability of temperature increases have not yet been studied.

B. Validity of BHTE

The BHTE (Eq. 1) is one of several theoretical models that have been proposed for heat transfer in vascularized tissues, all of which are approximations (Baish, 2000; Hristov, 2019). The BHTE was initially formulated under the assumption that heat exchange occurs in the capillary bed. In fact, most heat exchange in tissue occurs in larger vessels of diameter ranging from 80 μm to 1 mm (Baish, 2000). Smaller vessels, e.g. capillaries, are thermally equilibrated with surrounding tissue and do not transport heat through tissue, while larger vessels are too few in number to transport a significant amount of heat in tissue (but they do create temperature gradients in their vicinity and set up different heat transfer processes such as counter current heat flow). While such vessels can be included explicitly in a thermal model for RF dosimetry (e.g. Kotte *et al.* (1996) and Flyckt *et al.* (2007)), that would greatly complicate the problem and may require more data than is available.

While this does not invalidate the use of the BHTE, it points to the need for caution in its interpretation. In the BHTE, the "temperature" has to be interpreted as an average over a control volume that encompasses many thermally significant vessels, and which is not close to larger blood vessels. The blood flow parameter in the BHTE is conceptually different from tissue perfusion that is measured by laser Doppler flowmetry, for example (which measures velocity of red blood cells as opposed to volumetric flow). As Baish (2000) pointed out, "*As long as $\omega_b$ [the blood perfusion parameter] and $T_a$ [arterial temperature, identified as $T_b$ in Eq. 1] are taken as adjustable, curve-fitting parameters rather than literally as the perfusion rate and arterial blood temperature, the model may be used fruitfully, provided that the results are interpreted accordingly.*" For an example of a successful use of curve-fitting the blood perfusion parameter for hyperthermia treatment planning, see (Verhaart *et al.*, 2014; DeFord *et al.*, 1990). By contrast, nearly all of the thermal modelling studies discussed above were based on blood perfusion and other parameters taken from the literature, which will provide a representative value of the temperature increase in an actual subject but the results are hardly exact.

In spite of these caveats, limited comparisons with data show that the BHTE does an excellent job of predicting transient temperature increases, and a reasonable job of predicting long-term (steady state) temperature increases in RF-exposed tissues over a wide range of exposures (Kanezaki *et al.*, 2010; Foster *et al.*, 2018a). No studies, however, have been reported in which a representative group of human subjects were exposed to RF energy in the presently considered frequency range with measurements of the resulting increase in skin temperature. Nearly all thermal modeling studies have employed standard literature values for thermal parameters, few have experimentally validated the model results, and only a few studies have used subject-specific values of blood flow (Laakso *et al.*, 2017b). Uncertainties could be partly assessed through the use of a Monte Carlo analysis to calculate a distribution of temperatures over different combinations of parameters (e.g. Sasaki *et al.* (2017); (Li *et al.*, 2019b)) but ultimately more data are needed.

### C. Thermoregulation

Except one study (Kodera *et al.*, 2018a), thermoregulation was not considered in the computational evaluation of the local temperature rise. Skin blood flow is controlled by both core body and local skin temperature. Raising skin temperature from 32 °C to 40 °C at an ambient temperature of 22 °C results in a >10-fold increase in skin blood flow (Song *et al.*, 1989; Charkoudian, 2003). (However, local cutaneous microvascular responsiveness is impaired in patients with type 2 diabetes mellitus, making them more susceptible to heat stress (Charkoudian, 2003)). A thermal model that does not take into account the rapid increase in skin blood flow with skin temperature will overpredict temperature increases in skin, increasing safety margin. Conversely, cool ambient temperatures will lead to lower skin temperatures and reduce skin blood flow (Milan (1961)) and will cause a model to under-predict the RF-induced rise in skin temperature, although such errors are probably of secondary importance when designing RF safety limits. However, the effects of thermoregulatory responses on RF-induced increases in skin temperature have not yet been well studied and remain to be clarified.

*D. Whole-body Exposures*

The IEEE C95.1 standard and ICNIRP guideline have extended the frequency range of exposure reference level (IEEE) and reference level and basic restrictions (ICNIRP up to 300 GHz, vs. 100 GHz in previous versions of the guidelines. Limits on whole-body exposure are designed to take into account the total heat load on the human body from exposure at these frequencies. Only two studies considered here have reported changes in core body temperature from whole body exposures, and only up to 6 GHz (Hirata *et al.*, 2013; Moore *et al.*, 2017).

As in comments in Sec. 5C, most of the computational studies reported here assume an ambient (or environmental) temperature of around 22 – 28 °C (room temperature). In a modeling study, Moore *et al.* (2017) investigated the effects of exposure to RF energy in environments with elevated temperatures and high relative humidity, considering situations where heavy protective clothing must be worn. One scenario assumed exposure to RF energy at 6 GHz in an environment with ambient temperature of 38 °C and relative humidity of 60% (nearly an intolerable environment, with a 'heat index' value on the threshold of 'extreme danger'). In this environment, whole body exposure at 252.5 W/m$^2$ (whole body average SAR of 0.4 W/kg) increased local temperature by only 0.3 °C in the eyes and the testes, with smaller and physiologically insignificant rises in skin, bone marrow, brain and core. These exposure levels are considerably higher than IEEE and ICNIRP occupational exposure limits (50 W/m$^2$). Clearly more work needs to be done in this area, although at this stage, it would appear that the effects of added RF will be minor in comparison to the effects of altered ambient conditions.

**6. Summary**

This paper reviewed the dosimetric/analytic studies for human exposure to radio frequencies above 6 GHz where new wireless communication systems were deployed. Systematic review has been conducted for the studies on steady-state temperature for sinusoidal wave exposures and transient temperature rise for short pulse or pulse-train exposures. Though a limited number of studies have been reported on experimental studies, fair agreement between analytical, computational, and experimental temperatures was observed. The research necessity, especially for experimental studies, has been outlined for quantifying the uncertainty of computational results as well as improving the rationale of the limits in the international guidelines/standards.

Appendix

A search strategy was developed to retrieved using Web of Science database covering the period from 1990 to 2020. All the retrieved papers are screened to assess soundness based on the title and abstract by two reviewers. Then, the full contents of the papers that passed the initial screening are revised and classified as 'relevant' or 'excluded' by one reviewer. Additional papers were screened in Google Scholar engine and included if they were of high relevance.

The papers were excluded on the basis that they were no computational dosimetry studies, review/commentary papers without any new results, or not relevant for dosimetries at frequencies above 6 GHz. In Secs. 4A and 5A, search strategies developed were summarized in Table A1 and A2, respectively. If the reviewers found technical weakness and limitation, points were mentioned.

**Table A1.** Search strategy (**skin and eye above 6 GHz**). The term 'relevant' indicates the number of studies identified from the database that were not excluded.

| Search data | TS=(Temperature$) | | | | | | |
|---|---|---|---|---|---|---|---|
| | AND TS=("Heating" OR "Thermal Model*" OR Bioheat OR "Bio-heat" OR "Power density") | | | | | | |
| | AND TS=("Millimet* Wave$" OR mmW$ OR "Millimet* Band$" OR "5G" OR "5th generation") | | | | | | |
| | AND TS=(eye$ OR skin OR Phantom$) | | | | | | |
| Identified from database | 56 | Excluded (not relevant) | 46 | Relevant | 10 | Identified from other sources | 4 |

|                          |                          |    |                        |    |          |   |                            |   |
|--------------------------|--------------------------|----|------------------------|----|----------|---|----------------------------|---|
| Included in analysis     | 14                       |    |                        |    |          |   |                            |   |

**Table A2.** Search strategy (**brief exposure**). The term 'relevant' indicates the number of studies identified from the database that were not excluded

| Search data | TS=(temperature$) |  |  |  |  |  |  |  |
|---|---|---|---|---|---|---|---|---|
|  | AND TS=("Heat* " OR "Thermal Model*" OR Bioheat OR "bio-heat") |  |  |  |  |  |  |  |
|  | AND TS=(Microwaves$ OR Millimet*  OR  mmW$ OR Radiofrequency OR "Radio-frequency") |  |  |  |  |  |  |  |
|  | AND TS=(Eye$ OR Skin OR Tissue$) |  |  |  |  |  |  |  |
|  | AND TS=(Brief OR "Short-Duration" OR Train OR "s Exposure$" OR "Time Constant$" OR "Few minutes" OR "Few seconds" OR "heating kinetics") |  |  |  |  |  |  |  |
| Identified from database | 43 | Excluded (not relevant) | 37 | Relevant | 6 | Identified from other sources | 3 |
| Included in analysis | 9 |  |  |  |  |  |  |  |